# Investigation of n-type dilute magnetic semiconductor property observed in amorphous AlNO alloy thin film incorporated with dilute nitrogen at 300K

Deena Nath[1, 4], U.P. Deshpande[2], N. V. Chandra Shekar[1, 3], Sujay Chakravarty[1*]

[1]*UGC-DAE Consortium for Scientific Research, Kalpakkam Node, Kokilamedu-603104, India*

[2]*UGC- DAE Consortium for Scientific Research, Khandwa Road, Indore-452001, India*

[3]*Material Science Group, IGCAR, Kalpakkam-603102, India.*

[4]*University of Madras, Chennai, Tamil Nadu – 600005, India*

***Corresponding Author**: scha@csr.res.in

**Abstract** In the present work, a thin film was deposited on quartz substrate by reactive RF magnetron sputtering of high purity (99.999%) aluminium target using ultra-high pure (Ar + $N_2$) gas mixture. The percentage ratio of Ar and $N_2$ in the gas mixture was 95% and 5%, respectively. Chemical characterization using x-ray photoelectron spectroscopy (XPS) and energy-dispersive x-ray (EDX) spectroscopy reveals that in the presence of dilute nitrogen, Al prefers to react with residual oxygen to form $Al_2O_3$ while the nitrogen is incorporated in it. The stoichiometry of bulk film is $Al_2N_{0.38}O_{3.1}$. Magnetic and electrical properties measurement shows that the film exhibits n-type dilute magnetic semiconductor (DMS) property at 300K. The film has low electrical resistivity of 6.3 Ω-cm and high carrier mobility of $5.7\times10^6$ $cm^2V^{-1}s^{-1}$ at 300K. A density functional theory (DFT) calculation was performed to investigate the origin of observed magnetism in the film. From first-principles calculation based on DFT, it is found that for thermodynamic stability dilute nitrogen incorporated in $Al_2O_3$ preferred to sit at the interstitial site, which is responsible for observed magnetic property. Present study reported here provides a new insight to prepare rarely observed n-type DMS at room temperature by incorporating nitrogen interstitials in $Al_2O_3$, which is desirable for potential application in the field of spintronics.

## 1. Introduction-

For the emerging field of spintronics, it is essential to develop semiconductors with ferromagnetically polarized carriers at room temperature (RT) such that the spin along with charge of the carriers can be coupled with an external magnetic field to control functionalities of devices.[1] Dilute magnetic semiconductors (DMS) are one of the promising candidates to the emerging field of spintronics.

Among many, one of the desirable candidates for application as DMS is aluminium nitride (AlN). Few studies have been carried out to investigate RT ferromagnetism in AlN by incorporating magnetic or nonmagnetic elements [2,3]. However, no clear picture about origin of magnetism has emerged yet. Various possible reasons for the origin of ferromagnetism are reported in the literature, which includes aluminium vacancies as well as nitrogen vacancies. [4,5]. Furthermore, when AlN is deposited using reactive sputtering of the aluminium target with a mixture of Ar and $N_2$ gas, there is always a possibility of the presence of residual oxygen, which also reacts with the aluminium and results in the formation of AlNO alloy in the film. It is expected that AlNO alloy should have properties between pure AlN and $Al_2O_3$ i.e. exhibit high-temperature transmittance, high resistivity, variable refractive indices, and high bandgap energy depending upon oxygen and nitrogen content in the film [6]. The amount of oxygen present in the film depends on various factors like Ar:$N_2$ ratio, purity of Ar and $N_2$ gases, base vacuum, process pressure, etc. and in practice, it is almost impossible to get rid of oxygen entirely in the film. In literature, it is reported that the concentration of oxygen in the film decreases with increasing nitrogen ratio in Ar:$N_2$ gas mixture used for reactive sputtering during deposition [7]. Therefore, the film will consists of higher oxygen concentration when deposited in a dilute nitrogen environment. In literature systematic investigation has been carried out to understand the effect of oxygen on microstructure, chemical, optical and electronic properties of AlN deposited at higher nitrogen concentrations (≥50%) [7]. However, no such studies have been carried out on films deposited in dilute nitrogen environment (~ ≤10%). Furthermore, the effect of oxygen concentration on magnetic properties of AlN film has not been reported so far. Present work is motivated by the idea to study the effect of residual oxygen on microstructure, chemical, electronic and magnetic properties of AlN film during reactive sputter deposition in a dilute nitrogen environment.

## 2. Experimental details

A thin film was deposited at room temperature on quartz substrate by reactive RF magnetron sputtering of high purity (99.999%) aluminium target using ultra-high pure (Ar + $N_2$) gas mixture. The percentage ratio of Ar and $N_2$ in the gas mixture was 95% and 5%, respectively. The base pressure in the deposition chamber was $1.2 \times 10^{-6}$ mbar. The process pressure during the deposition of the film was kept at $3.5 \times 10^{-2}$ mbar. The substrate to target distance was fixed at 75 mm. The film was deposited for 60 min. The substrate was rotated at a speed of 10 rpm during deposition to obtain a homogeneous film. The RF power of the magnetron sputter gun was fixed at 200W during the deposition of the film. The thickness of the deposited film is 37 nm measured using a surface profilometer.

Microstructural characterization of the as-deposited film was carried out by Grazing Incident X-Ray Diffraction (GIXRD) measurement at a fixed incident angle of 0.3 degree using M/S Bruker D8 Discover instrument with 4.5 kW Cu $K_\alpha$ rotating anode as an X-ray source. To determine the surface chemical composition of the as-deposited film, X-ray photoelectron spectroscopy (XPS) measurement was carried out in commercial monochromatic XPS instrument (M/S SPECS Surface Nano Analysis GmbH, Germany). The XPS data has been recorded using a monochromatic X-ray source of aluminium with $K_\alpha$ = 1486.61 eV operated at 13 kV and 100W. XPS data had been acquired after pre sputtering the film surface for 5 min using 500 eV $Ar^+$ ions. The charge corrected XPS spectra after Shirley background subtraction were fitted with Gaussian Lorentzian line shapes. To determine the chemical composition of bulk thin film, EDX analysis has been carried out in Helios NanoLab™ 600i instrument (M/s FEI Quanta, Switzerland) equipped with an EDS detector from EDAX/AMETEK. Resistivity and Hall measurements were carried out using the four-probe method in a commercial 15T cryogen-free measurement system (CFMS 15) from M/s Cryogenic Ltd., UK. Four contacts were made in the film following Van der Pauw geometry using copper wire and highly conductive silver paste. Keithley DC source meter (model 2400) was used as a current source. Keithley nano voltmeter (model 2182A) was used to measure the corresponding drop in voltage. Resistivity measurement was carried out at fixed current $1\times10^{-7}$ A. Hall measurement was carried out at 300K using $1\times10^{-6}$ A fixed current and sweeping the applied magnetic field perpendicular to film surface between ±4T. Magnetic measurement of the film was carried out using commercial SQUID-VSM MPMS system from M/s Quantum Design, USA. For magnetic measurement, the magnetic moment (m) of the bare quartz substrate is first measured at 300K as a function of the applied magnetic field (H) by sweeping the field in the range between ± 5T. The magnetic field was applied parallel to the film surface. The measured m(H) graph of the bare quartz substrate exhibits a diamagnetic signal. After that film is deposited in the same quartz substrate and m(H) was measured again for the film plus substrate at 300K in similar steps used to measure bare quartz substrate. Finally, the m(H) of the bare film is obtained by subtracting the corresponding moment of the bare quartz substrate. In order to avoid magnetic contamination non-magnetic tweezer was used to handle the sample.

## 3. Results and Discussion

### 3.1 Structural characterization

Microstructural characterization of the as-deposited film carried out by GIXRD measurement is shown in **Fig. 1**. The GIXRD graph shows no clear Bragg peak indicating that the film is amorphous in nature.

### 3.2 Chemical characterization

To find the chemical composition close to the surface of the film XPS measurement was carried out on as-deposited film after 5 min pre-sputtering with the Ar ion. The fitted XPS spectra of Al 2p, N 1s, and O 1s are shown in **Fig. 2 (a), (b) and (c)**, respectively. N 1s spectrum is well fitted with three peaks corresponding to $N^{+1}$, $N^{+2}$, and $N^{+3}$ oxidation states. The absence of a peak corresponding to the $N^{-3}$ oxidation state ruled out the presence of AlN in the film. Similarly, both Al 2p and O 1s spectra get fitted using only two prominent peaks corresponding to $Al_2O_3$ and Al-O-N, which again confirms the absence of AlN in the film. The relative concentration of Al, N, and O at film surface was obtained using fitted peak area of Al 2p, N 1s and O 1s spectra, respectively after normalization with relative sensitivity factor (RSF) { $RSF_{Al2p}$ = 0.11, $RSF_{O1s}$ = 0.63, $RSF_{N1s}$ = 0.38} **[8]**. The Al, N, and O concentration obtained at the film surface is approximately 41%, 1%, and 58% respectively. Hence, the film surface has a chemical stoichiometry of $Al_2N_{0.05}O_{2.8}$, which is very close to $Al_2O_3$ with dilute nitrogen incorporated in it. Furthermore, Energy Dispersive X-Ray [EDX] analysis was carried out to determine the composition of the bulk thin film, which is shown in **Fig. 2(d)**. The relative concentration of Al, N & O in bulk film obtained from EDX is approximately 37%, 7%, and 57% respectively and hence the stoichiometry of bulk film is $Al_2N_{0.38}O_{3.1}$, which is also very close to the stoichiometry of the film at the surface obtained from XPS. Thus, from XPS and EDX analysis it is concluded that the whole film consists of $Al_2O_3$ with dilute nitrogen incorporated in it. The amount of nitrogen incorporated in $Al_2O_3$ in the bulk film is higher (~7%) than that at close to the surface (~1%).

In an earlier microstructure study of $AlN_xO_y$ thin films, it was observed that when the concentration ratio of non-metallic to metallic elements ($C_{N+O}/C_{Al}$) is equal to 0.85 or more, the system becomes amorphous [9]. In the present case, the ratio of $C_{N+O}/C_{Al}$ in the film varies between 1.4 to 1.7 from the surface to bulk, which explains the reason for the amorphous nature of film confirmed from GIXRD measurement.

*3.3 Electrical transport measurement*

The sheet resistance of the film was measured as a function of temperature (T) using the Van der Pauw four-probe method [10]. The sheet resistivity $\rho_s$ was calculated using the following expression

$$R_s = \frac{\pi}{ln(2)} \times \frac{(R'+R'')}{2} \times f\left(\frac{R'}{R''}\right) \tag{1}$$

$$\rho_s = R_s \times t \tag{2}$$

Where $R_S$ is the sheet resistance, $R'$ & $R''$ is resistance measured in horizontal and vertical direction respectively, $f(R'/R'')$ is Van der Pauw correction factor and t is the thickness of the deposited film. As shown in **Fig. 3(a)** resistivity of the film increases with decreasing temperature indicating the semiconducting nature. At room temperature (300K), the film exhibits low electrical resistivity of 6.3 Ω-cm [9]. Furthermore, hall measurement was carried out at 300K. The variation in hall voltage ($V_H$) as a function of the external applied field ($B$) is shown in **Fig. 3(b)**. The slope between $V_H$ & $B$ is negative indicating electrons as majority carriers and film behaves as n-type semiconductors. The majority carrier concentration was calculated from the slope using the following expression:

$$n_s = \frac{IB}{qV_H} \tag{3}$$

Where $I$ is applied current and $q$ is charge of the electron. The majority carrier concentration in the film at 300K calculated using **equation (3)** is $(1.7 \pm 0.04) \times 10^{11}$ cm$^{-3}$, which is several orders of magnitude less than that observed in the conventional semiconductors. The low carrier concentration can be attributed due to dilute nitrogen sites incorporated in $Al_2O_3$ responsible for n-type majority carriers.

The mobility ($\mu$) of the majority carriers at 300K is calculated using the following expression.

$$\mu = \frac{\gamma}{q \times n_s} \tag{4}$$

Where $\gamma$ is the conductivity of thin-film ($\gamma = 1/\rho_s$), $q$ is a charge of the electron and $n_s$ is the concentration of majority carrier. The majority carrier mobility in the film at 300K calculated using **equation (4)** is $5.7 \times 10^6$ cm$^2$V$^{-1}$s$^{-1}$, which is very high.

In summary, from transport measurement it is concluded that film behaves as n-type semiconductor exhibiting low resistivity and very high carrier mobility at 300K.

*3.4 Magnetic measurement*

The magnetic moment ($m$) of the film as a function of the applied magnetic field ($H$) was measured at 300K. In **Fig. 4(a)** the $m(H)$ data plot for the bare film is shown after subtracting the contribution from the quartz substrate, which exhibits ferromagnetic hysteresis loop superimposed on a paramagnetic signal. Measurement of *m(H)* on the bare quartz substrate was carried out before film deposition, which is also shown in **Fig. 4(a)**. For clarity, zoomed m(H) plot at low field region is shown in inset .

Contribution of the ferromagnetic hysteresis loop (FM) was separated from the paramagnetic (PM) part by fitting the experimentally measured *m(H)* data plot using the following **equation (5) [11]**:

$$M(H) = \left[2\frac{M_{FM}^S}{\pi} tan^{-1}\left\{\frac{(H \pm H_{ci})}{H_{ci}} tan\left(\frac{\pi M_{FM}^R}{2M_{FM}^S}\right)\right\}\right] + \chi H \qquad (5)$$

Where $M^S_{FM}$ is saturation magnetization, $H_{ci}$ is intrinsic coercivity and $M^R_{FM}$ is remanence of the FM part, while $\chi$ is the susceptibility of PM part and *M(H)* is the magnetization of the film measured as a function of the applied magnetic field. The magnetic moment (*m*) was converted into magnetization (*M*) by dividing it with film volume. The fitted M(H) graph is shown in **Fig. 4(b)** and its inset show the fitting quality at low magnetic field. The fitted parameters are tabulated in **Table 1**. It can be seen from **Table 1** that the susceptibility $\chi$ of the PM part is $6.9 \times 10^{-2}$. The saturation magnetization ($M^S_{FM}$) and remanence ($M^R_{FM}$) of FM part is $1.14 \times 10^4$ JT$^{-1}$m$^{-3}$ ($\approx$ 11emu/cc) and $7 \times 10^2$ JT$^{-1}$m$^{-3}$ ($\approx$ 0.7 emu/cc), respectively. The important point that can be observed in **Table 1** is that the intrinsic coercivity ($H_{ci}$) and Stoner-Wohlfarth number ($M^S_{FM} / M^R_{FM}$) of FM part is (49 ± 6) Oe and (0.07 ± 0.007) respectively, which is very small. Therefore, the film can be considered consisting of non-interacting superparamagnetic (SPM) particles, which contributes to the FM part. Following Chantrell et al. **[12,13]** for dispersion of non-interacting SPM particles throughout the film, the magnetization can be expressed as,

$$M(H) = M_s \int_0^\infty L\left[\frac{\mu H}{k_B T}\right] F_V(\mu)\, d\mu = M_s \int_0^\infty L\left[\frac{M_{sb}V_{mv}H}{k_B T}y\right] F_V(y)\, dy \qquad (6)$$

Where $L\left[\frac{\mu H}{k_B T}\right]$ is the Langevin function, $M_s$ is the saturation magnetization of the system, $\mu$ (=$M_{SB}V$) is the particle magnetic moment, $M_{SB}$ is the bulk saturation magnetization and $V_{mV}$ is the particles median volume in the volume-weighted distribution. $F_V(\mu)$ represents the volume-weighted distribution of the moment while $F_V(y)$ represents the volume-weighted distributions of the reduced particle volumes *(V/V$_{mV}$)*, where *V* is the total volume of the film. To fit the *M(H)* virgin curve measured at 300K *(RT M-H)*, **equation (6)** has been modified by introducing paramagnetic term as expressed in Eq. (7) below.

$$M(H) = M_s \int_0^\infty L\left[\frac{\mu H}{k_B T}\right] F_V(\mu)\, d\mu + \chi H = M_s \int_0^\infty L\left[\frac{M_{sb}V_{mv}H}{k_B T}y\right] F_V(y)\, dy + \chi H \qquad (7)$$

It is reported that in the case of SPM particles it is appropriate to use lognormal distribution (LND) as given below **[12,13]**,

$$f(x)dx = \frac{exp\left(-\frac{[ln(\frac{x}{x_m})]^2}{2\sigma_x^2}\right)}{\sqrt{2\pi}\,\sigma_x x}\, dx \qquad (8)$$

Where x$_m$ is the median and $\sigma_x$ is the standard deviation of *ln(x)*.

The first quadrant of the RT-MH graph fitted using **equation (7)** is shown in **Fig. 4(c)**. The adj.$R^2$ which defines the quality of fitting was 0.99998 (close to 1) showing a very good fit. The parameters obtained after fitting is tabulated in **Table 2**. It can be seen in **Table 2** that the bulk saturation magnetization ($M_{SB}$) of SPM particles contributing to the FM part is $21.1\times10^4$ JT$^{-1}$m$^{-3}$ ($\approx$ 211 emu/cc). However, the magnetic volume packing fraction ($\varepsilon$) obtained by ($M_S/M_{SB}$) is only 0.057, which indicates that the SPM particles occupy only 5.7% of total film volume. The total magnetic volume ($V_{Mag}$) of SPM particles is $1.85\times10^{13}$ nm$^3$, while the total film volume is $32.6\times10^{13}$ nm$^3$. Furthermore, the LND plot of the number of SPM particles as a function of particle volume obtained using **equation (8)** is shown in **Fig. 4(d)**. The mode of the LND plot is 45 nm$^3$. The median and standard deviation of the LND plot is also given in **Table 2**, which is 667 nm$^3$ and 1.2, respectively. The LND plot is highly asymmetric and there is a large deviation observed between median and mode value indicating random volume distribution of SPM particles in the film.

The magnetic measurement results described above confirms the presence of magnetic moment in the film at 300K. In the majority of the film, the magnetic moments are isolated and non-interacting contributing to the PM part. However, in a small volume fraction of the film (~ 5.7%), the magnetic moments are close enough to interact with each other in a short-range (~ volume 45 nm$^3$) forming non-interacting SPM particles, which contributes to FM part.

### 4. Theoretical simulation

The next question, which needs to be addressed, is the origin of the observed magnetic moment in the film. From XPS and EDX results it is observed that the stoichiometry of the film consists of $Al_2O_3$ with dilute nitrogen incorporated in it. Hence, the following two possibilities can be considered responsible for an observed magnetic moment in the film: 1) presence of pristine $Al_2O_3$ and 2) the presence of one extra nitrogen sitting at the interstitial site in $Al_2O_3$ i.e., $(Al_2O_3)_{1N}$.

In order to further investigate, the density of state (DOS) and band structure was theoretically simulated using first-principles calculations based on DFT on supercell (space group R3c[167]) of size $2\times1\times1$ of $Al_2O_3$ and $(Al_2O_3)_{1N}$, respectively. A small supercell has been chosen for computation since the deposited film is amorphous (confirmed from GIXRD) indicating the absence of long-range ordering. The spin polarised calculation is performed using the CASTEP (Cambridge Sequential Total Energy Package) program code in Materials Studio software (version 7.0) [14]. Generalized gradient approximation (GGA) of the Perdew-Burke Ernzerhof (PBE) form was used as an exchange-correlation function [15]. Self-consistent field calculation (Tolerance - $5.0\times10^{-7}$ eV/atom) was employed in a plane wave basis. The norm-conserving pseudopotential is used to describe the electron-ion interaction with cut-off energy of 830 eV. A ($3\times7\times7$) Monkhorst- pack mesh is used for the generation of the Brillouin zone during geometry optimization in the supercell. The structure is fully relaxed by geometrical optimization calculation by the motion of the atomic position and lattice parameter with the convergence threshold of the remanent Hellmann-Feynman force of 0.001 eV/Å. Finally, on the relaxed structure, the density of state (DOS) and bandgap calculation were performed. Furthermore, the spin-polarized molecular dynamics simulation was carried out on optimized ($2\times1\times1$) supercell of $(Al_2O_3)_{1N}$.

The optimized supercell of $Al_2O_3$ and $(Al_2O_3)_{1N}$ used for calculation is shown in **Fig. 5 (a) and (b)**, respectively. The ground state energy obtained for the optimized supercell of $Al_2O_3$ and $(Al_2O_3)_{1N}$ is -5673.12 eV & -5935.89 eV, respectively. It can be noticed that the ground state energy of $(Al_2O_3)_{1N}$ is lower than $Al_2O_3$ indicating that $(Al_2O_3)_{1N}$ is thermodynamically more stable than $Al_2O_3$. Therefore, the nitrogen incorporated in $Al_2O_3$ will prefer to sit at the interstitial site.

Simulated spin-resolved DOS and band structure along high symmetry point in the Brillouin zone for optimized supercell of $Al_2O_3$ is shown in **Fig. 5 (c) and (d)**, respectively. In the case of $Al_2O_3$ large direct bandgap of 7.06 eV is observed, which is very close to that reported in the literature **[16]**. In **Fig. 5(c)** it can be observed that there is symmetry in both spins up and spin down states and hence the net magnetic moment for $Al_2O_3$ is 0 $\mu_B$. Thus, $Al_2O_3$ is non-magnetic and could not contribute to the magnetic moment observed in the film.

Simulated spin-resolved DOS and band structure for optimized supercell of $(Al_2O_3)_{1N}$ are shown in **Fig. 5 (e) and (f)**, respectively. As can be seen in **Fig. 5(e)**, there is an asymmetry in spin up and spin down states with a net magnetic moment of 1 $\mu_B$. Therefore, it is concluded that $(Al_2O_3)_{1N}$ mainly contributes to the magnetic moment observed in the film. Furthermore, from **Fig. 5(c)** and **Fig. 5(e)** it can be noticed that in case of $(Al_2O_3)_{1N}$, i.e., inserting one nitrogen at the interstitial site in $Al_2O_3$ results in the creation of some new spin-down states near to the Fermi level, which is absent in $Al_2O_3$. Hence, there is also a possibility of spin-polarized transport in $(Al_2O_3)_{1N}$. The simulated band structure of optimized $(Al_2O_3)_{1N}$ supercell has an indirect bandgap of 1.72 eV and 4.1 eV for spin down and spins up sates, respectively as illustrated in **Fig. 5(f)**. The indirect bandgap observed in $(Al_2O_3)_{1N}$ might be also useful for application in the field of optoelectronics and photonics and needs to be explored further, which is beyond the scope of the present work.

Spin-polarized molecular dynamics (MD) simulation has been also carried out on an optimized $(Al_2O_3)_{1N}$ supercell to examine whether the net magnetic moment also sustains at RT. The molecular dynamics simulation is performed with Nose-Hoover thermostat at 300K and the simulation time was kept as 0.5 ps. The fluctuation in the temperature during simulation steps and the corresponding changes in the magnetic moment are shown in **Fig. 6 (a) and (b)**, respectively. The system sustains an average magnetic moment of ~ 1$\mu_B$ at 300K during the simulation time of 0.5 ps. The structure remains unaffected after MD simulation due to the larger binding energy of Al-Al and Al-O bonds compared to thermal energy at 300K. Thus, MD simulation confirms that $(Al_2O_3)_{1N}$ sustains a magnetic moment at room temperature.

In summary, from theoretical simulation, it is confirmed that for thermodynamic stability the dilute nitrogen incorporated in $Al_2O_3$ prefers to sit at the interstitial sites. The nitrogen interstitial sites in $Al_2O_3$ induce a net magnetic moment, which sustains at 300K.

## 5. Conclusions

Residual oxygen plays a significant role during the deposition of AlN film using reactive sputtering of the aluminium target in a dilute nitrogen environment. In the presence of dilute nitrogen Al prefers to react with residual oxygen to form $Al_2O_3$ while the nitrogen is incorporated in it. For thermodynamic stability, nitrogen prefers to sit at interstitial sites in $Al_2O_3$. The nitrogen interstitial sites at $Al_2O_3$ induce magnetic moment in the film, which sustains at 300K. The nitrogen interstitial sites are also responsible for n-type majority carriers. Furthermore, inserting one nitrogen at the interstitial site in $Al_2O_3$ results in the creation of some new spin-down states near to the Fermi level, which is absent in $Al_2O_3$. Hence, there is also a possibility of spin-polarized transport in $(Al_2O_3)_{1N}$. The film exhibits rarely observed n-type Dilute Magnetic Semiconductor (DMS) property at room temperature (RT) desirable for potential application in the field of spintronics.

**Acknowledgments**

Authors would like to thank Dr. Shamima Hussain & Siddhartha Dam of UGC DAE CSR, Kalpakkam Node for their help in EDX measurement. Authors would also like to thank Balaram Thakur of UGC DAE CSR, Kalpakkam Node for his help in a theoretical simulation study and fruitful discussion.

**References**

[1] G. Yao, G. Fan, H. Xing, S. Zheng, J. Ma, Y. Zhang, L. He, Electronic structure and magnetism of V-doped AlN, J. Magn. Magn. Mater. 331 (2013) 117–121. https://doi.org/10.1016/j.jmmm.2012.11.031.

[2] S.G. Yang, A.B. Pakhomov, S.T. Hung, C.Y. Wong, Room-temperature magnetism in Cr-doped AlN semiconductor films, Appl. Phys. Lett. 81 (2002) 2418–2420. https://doi.org/10.1063/1.1509475.

[3] D. Pan, J.K. Jian, A. Ablat, J. Li, Y.F. Sun, R. Wu, Structure and magnetic properties of Ni-doped AlN films, J. Appl. Phys. 112 (2012). https://doi.org/10.1063/1.4749408.

[4] H.H. Ren, R. Wu, J.K. Jian, C. Chen, A. Ablat, Al vacancy induced room-temperature ferromagnetic in un-doped ALN, Adv. Mater. Res. 772 (2013) 57–61. https://doi.org/10.4028/www.scientific.net/AMR.772.57.

[5] Y. Liu, L. Jiang, G. Wang, S. Zuo, W. Wang, X. Chen, Adjustable nitrogen-vacancy induced magnetism in AlN, Appl. Phys. Lett. 100 (2012) 2012–2015. https://doi.org/10.1063/1.3696023.

[6] P.W. Wang, J.C. Hsu, Y.H. Lin, H.L. Chen, Structural investigation of high-transmittance aluminum oxynitride films deposited by ion beam sputtering, Surf. Interface Anal. 43 (2011) 1089–1094. https://doi.org/10.1002/sia.3700.

[7] M.A. Signore, A. Taurino, D. Valerini, A. Rizzo, I. Farella, M. Catalano, F. Quaranta, P. Siciliano, Role of oxygen contaminant on the physical properties of sputtered AlN thin films, J. Alloys Compd. 649 (2015) 1267–1272. https://doi.org/10.1016/j.jallcom.2015.05.289.

[8] P.W. Wang, J.C. Hsu, Y.H. Lin, H.L. Chen, Nitrogen bonding in aluminum oxynitride films, Appl. Surf. Sci. 256 (2010) 4211–4214. https://doi.org/10.1016/j.apsusc.2010.02.004.

[9] J. Borges, N. Martin, N.P. Barradas, E. Alves, D. Eyidi, M.F. Beaufort, J.P. Riviere, F. Vaz, L. Marques, Electrical properties of AlN xO y thin films prepared by reactive magnetron sputtering, Thin Solid Films. 520 (2012) 6709–6717. https://doi.org/10.1016/j.tsf.2012.06.062.

[10] F. Werner, Hall measurements on low-mobility thin films, J. Appl. Phys. 122 (2017). https://doi.org/10.1063/1.4990470.

[11] S.R. Mohapatra, P.N. Vishwakarma, S.D. Kaushik, R.J. Choudhary, N. Mohapatra, A.K. Singh, Cobalt substitution induced magnetodielectric enhancement in multiferroic Bi2Fe4O9, J. Appl. Phys. 121 (2017) 0–11. https://doi.org/10.1063/1.4979094.

[12] R.W. Chantrell, J. Popplewell, S.W. Charles, Measurements of particle size distribution parameters in ferrofluids, IEEE Trans. Magn. 14 (1978) 975–977. https://doi.org/10.1109/TMAG.1978.1059918.

[13] M. El-Hilo, Nano-particle magnetism with a dispersion of particle sizes, J. Appl. Phys. 112 (2012). https://doi.org/10.1063/1.4766817.

[14] M.D. Segall, P.J.D. Lindan, M.J. Probert, C.J. Pickard, P.J. Hasnip, S.J. Clark, M.C. Payne, First-principles simulation: Ideas, illustrations and the CASTEP code, J. Phys. Condens. Matter. 14 (2002) 2717–2744. https://doi.org/10.1088/0953-8984/14/11/301.

[15] J.P. Perdew, K. Burke, M. Ernzerhof, Generalized gradient approximation made simple, Phys. Rev. Lett. 77 (1996) 3865–3868. https://doi.org/10.1103/PhysRevLett.77.3865.

[16] E.O. Filatova, A.S. Konashuk, Interpretation of the Changing the Band Gap of $Al_2O_3$ Depending on Its Crystalline Form: Connection with Different Local Symmetries, J. Phys. Chem. C. 119 (2015) 20755–20761. https://doi.org/10.1021/acs.jpcc.5b06843.

## Table's caption

**Table 1** – Parameters obtained for FM and PM parts separately after fitting M(H) curve using equation (5).

**Table 2.** Parameters obtained after fitting the first quadrant of the M-H curve at 300K using equation (7).

## Figure's caption

**Fig. 1** – GIXRD spectra of as deposited film measured at a fixed incident angle of 0.3 degree.

**Fig. 2** – Fitted XPS spectra of *a)* Al 2p, *b)* N 1s and *c)* O 1s binding energy peaks after charge correction and *d)* EDX spectra from bulk thin film

**Fig. 3** – *a)* DC resistivity as a function of temperature. The resistivity increases with decreasing temperature. *b)* Hall voltage as a function of the externally applied field. The slope of the plot is negative.

**Fig. 4** – *a)* m(H) curve measured at 300K for bare quartz substrate and after deposition of a film on the quartz substrate. The m(H) curve of the bare film obtained after subtracting the contribution from the quartz substrate is also shown. For clarity, zoomed m(H) plot at low field region is shown in inset. *b)* Fitted M(H) hysteresis loop of bare film at 300K using equation (5) to separate the contribution of FM part from PM part. Inset shows the fitting quality at low magnetic field *c)* Fitted first quadrant of measured M(H) curve at 300K using equation (7). *d)* LND plot of the number of SPM particles as a function of particle volume obtained using equation (8).

**Fig. 5** – Theoretically simulated optimized structure of *a)* $Al_2O_3$ and *b)* $(Al_2O_3)_{1N}$ supercell obtained using first principle calculation based on the density functional theory. Pink, red & blue balls represent aluminium, oxygen, and nitrogen atoms respectively. Calculated *c)* DOS and *d)* band structure of optimized $Al_2O_3$ supercell. Calculated *e)* DOS and *f)* band structure of optimized $(Al_2O_3)_{1N}$ supercell

**Fig. 6** –Fluctuation in *a)* temperature and *b)* magnetic moment observed in molecular dynamics simulation carried out at 300K on optimized $(Al_2O_3)_{1N}$ supercell.

**Table 1** – Parameters obtained for FM and PM parts separately after fitting M(H) curve using equation (5).

| Saturation magnetization | Remanence magnetization | Coercivity | Stoner-Wohlfarth number or Squareness factor | Paramagnetic susceptibility |
|---|---|---|---|---|
| $M^S_{FM}$ ($*10^4$ J/T/m$^3$) | $M^R_{FM}$ ($*10^4$ J/T/m$^3$) | $H_{Ci}$ (O$_e$) | $K_p$ ($M^R_{FM}$ / $M^S_{FM}$) | $\chi$ ($*10^{-2}$) |
| 1.1 ± 0.03 | 0.07 ± 0.007 | 49 ± 6 | 0.07 ± 0.007 | 6.9 ± 0.003 |

**Table 2.** Parameters obtained after fitting the first quadrant of the M-H curve at 300K using equation (7).

| Saturation magnetisation | Bulk saturation magnetisation | Median volume of magnetic particle | Standard deviation | Magnetic volume packing fraction | Film volume | Magnetic volume $V_{Mag} = \varepsilon * V_{Film}$ |
|---|---|---|---|---|---|---|
| $M_S$ ($*10^4$ J/T/m$^3$) | $M_{SB}$ ($*10^4$ J/T/m$^3$) | $V_{mV}$ (nm$^3$) | $\sigma$ | $\varepsilon$ $M_S / M_{SB}$ | $V_F$ ($*10^{13}$ nm$^3$) | $V_{Mag}$ ($*10^{13}$ nm$^3$) |
| 1.2 (± 0.03) | 21.1 (± 0.6) | 667 (± 18) | 1.2 (± 0.1) | 0.057 (± 0.002) | 32.6 | 1.85 (± 0.07) |

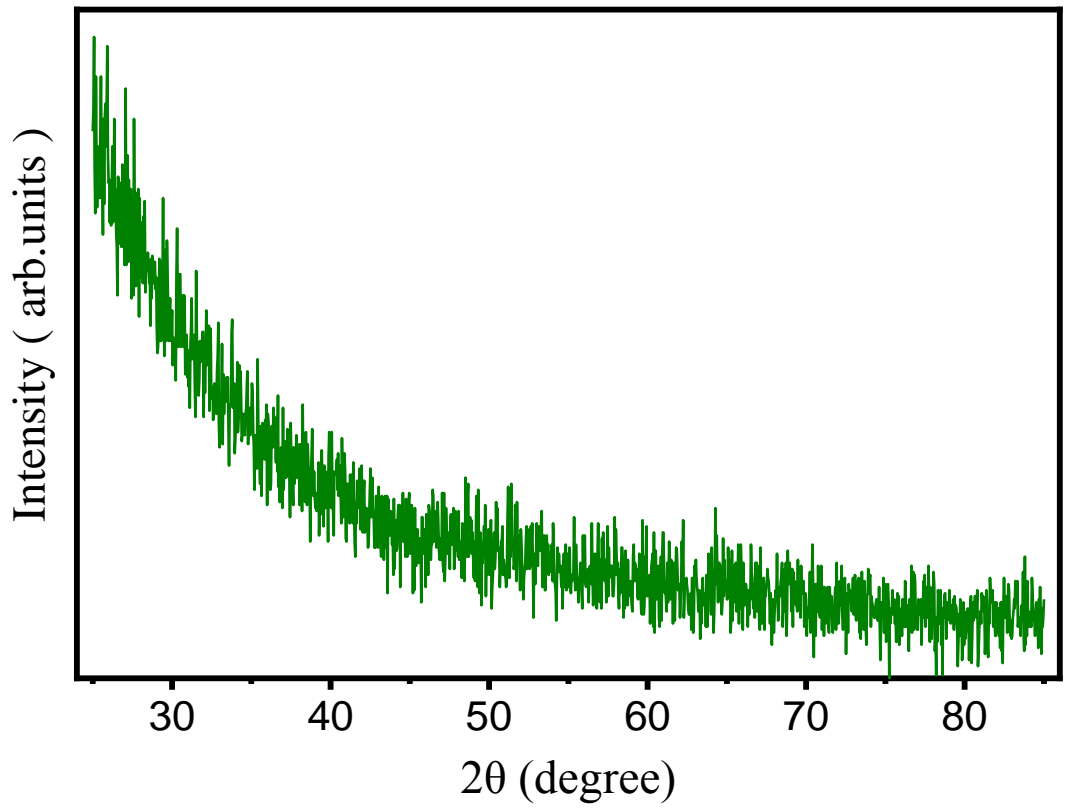

Fig. 1

**Fig. 2**

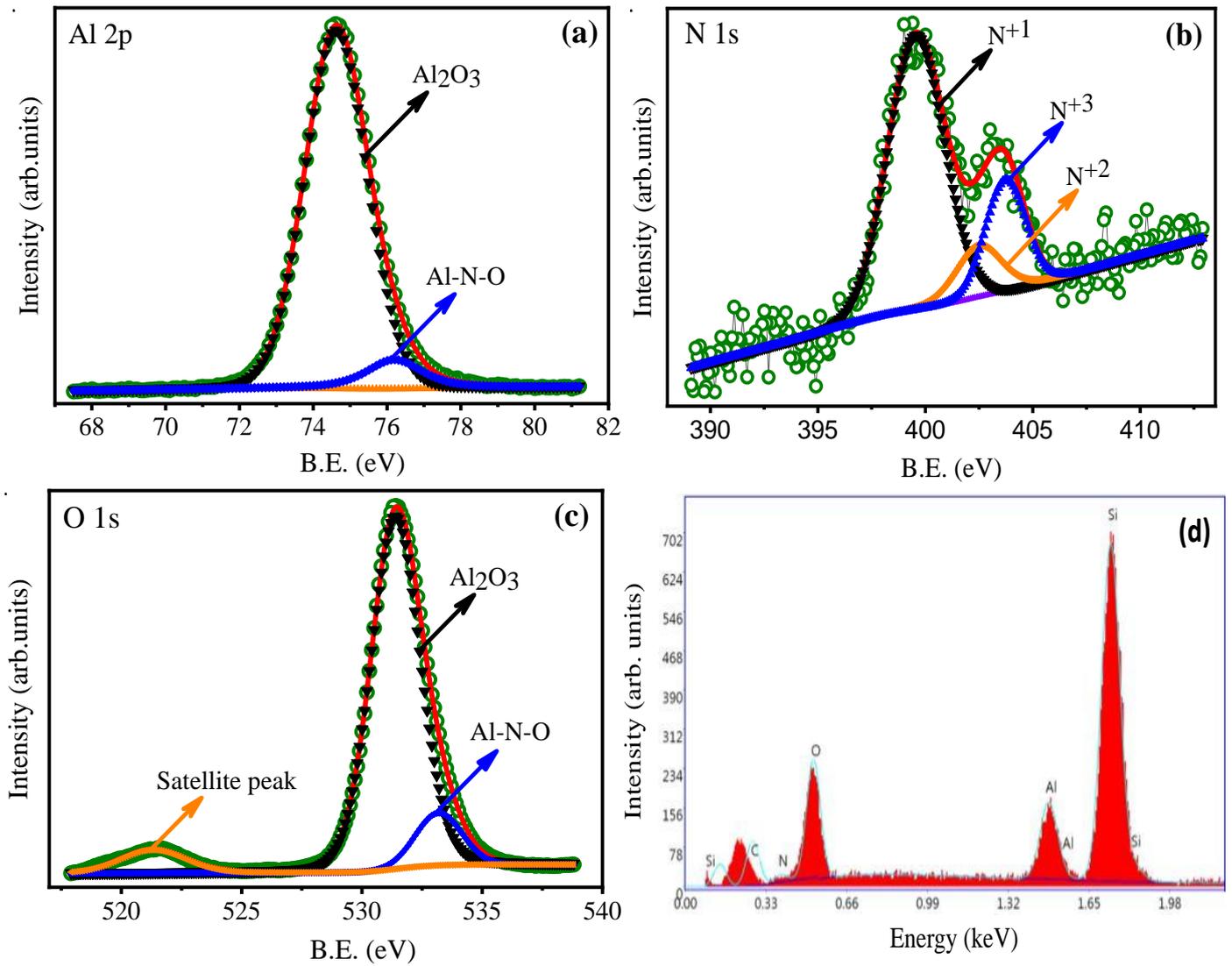

**Fig. 3**

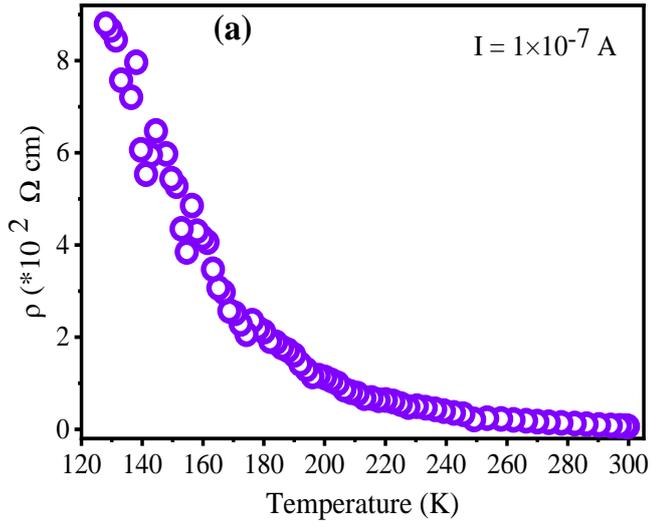 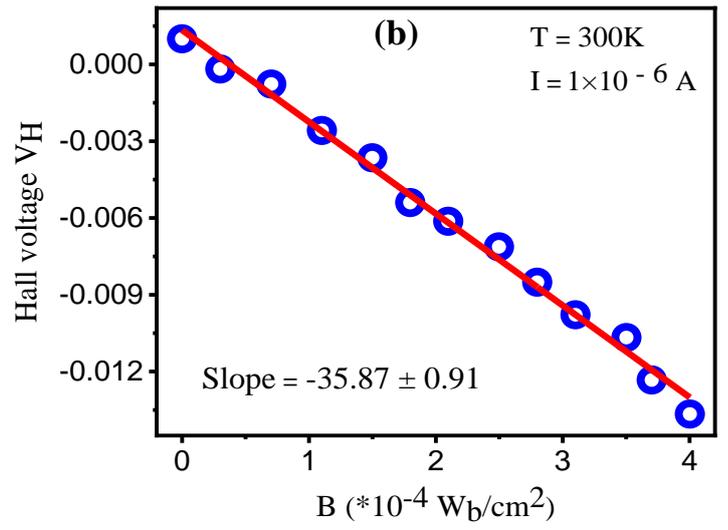

**Fig. 4**

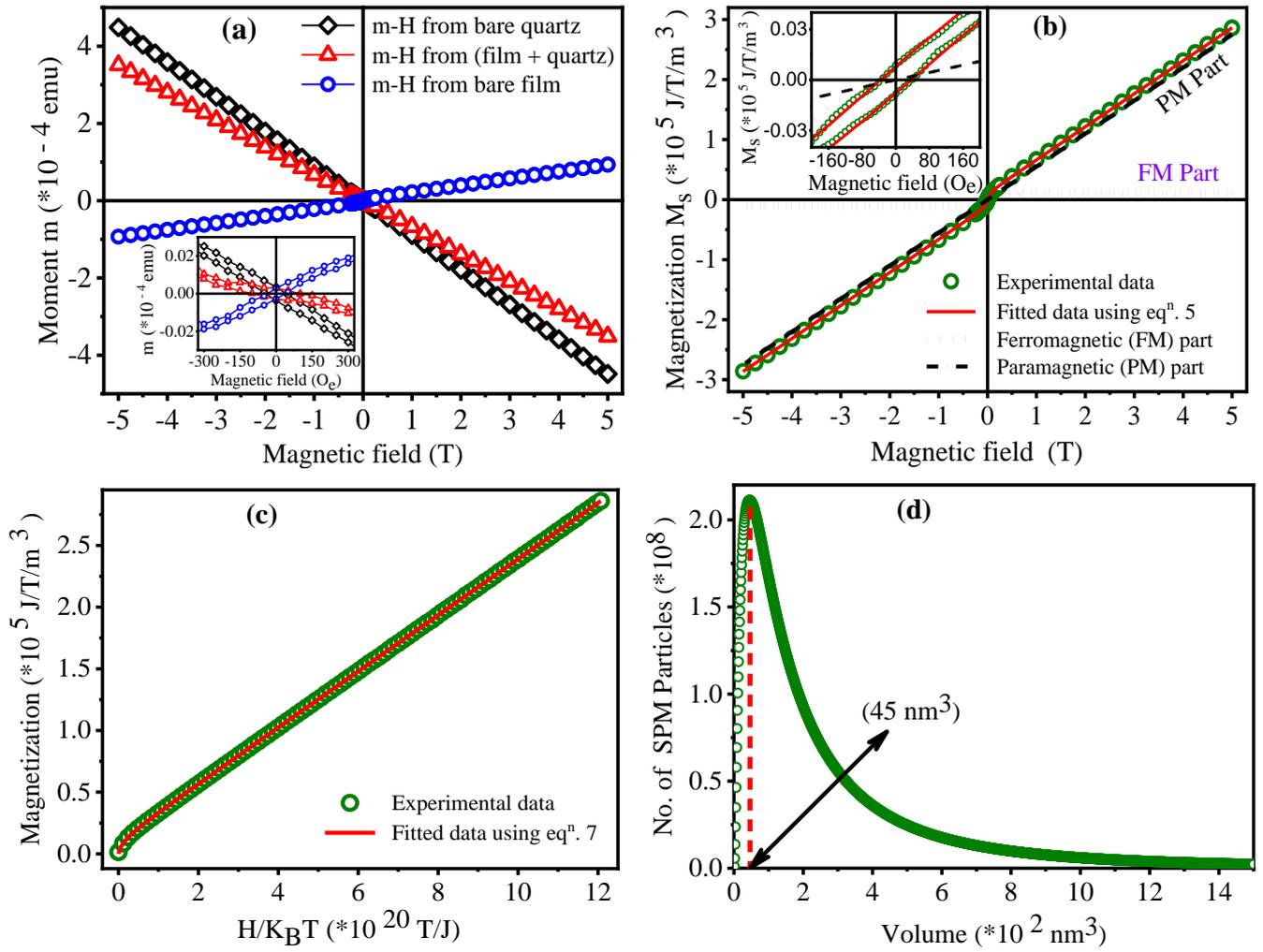

**Fig. 5**

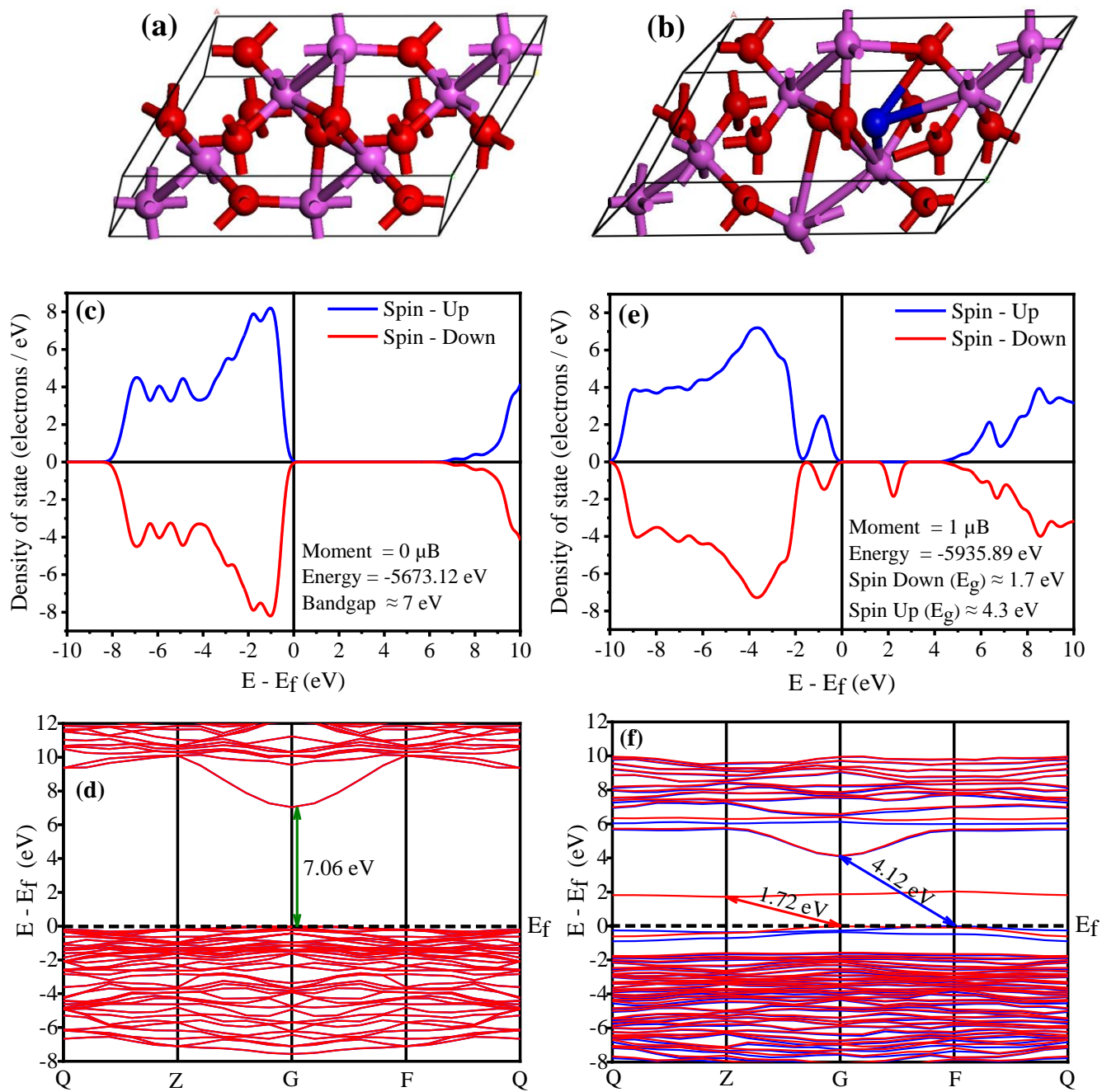

**Fig. 6**

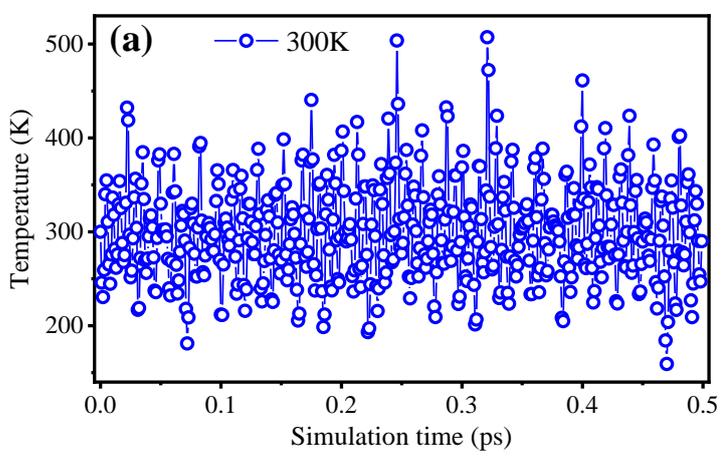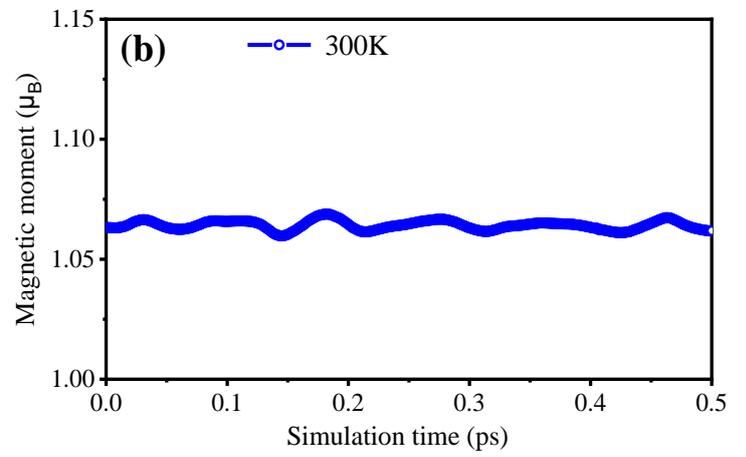